\documentclass[aip, jcp, twocolumn]{revtex4-1}
\usepackage{amsmath}
\usepackage{color}
\usepackage{hyperref}
\usepackage{graphicx}

\begin{document}
\title{Interference enhanced thermoelectricity in quinoid type structures}
\author{M. Strange}
\affiliation{Nano-Science center and Department of
Chemistry, University of Copenhagen, Universitetsparken 5, 2100 Copenhagen {\O}, Denmark}
\email{strange@chem.ku.dk}
\author{J. S. Seldenthuis}
\affiliation{Kavli Institute of Nanoscience, Delft University of Technology, 2628 CJ 
Delft, The Netherlands}
\author{C. J. O. Verzijl}
\affiliation{Kavli Institute of Nanoscience, Delft University of Technology, 2628 CJ 
Delft, The Netherlands}
\author{J. M. Thijssen}
\affiliation{Kavli Institute of Nanoscience, Delft University of Technology, 2628 CJ 
Delft, The Netherlands}
\author{G. C. Solomon}
\affiliation{Nano-Science center and Department of
Chemistry, University of Copenhagen, Universitetsparken 5, 2100 Copenhagen {\O}, Denmark}

\begin{abstract}
Quantum interference (QI) effects in molecular junctions may be used to obtain large thermoelectric responses. 
We study the electrical conductance $G$ and the thermoelectric response of a series of molecules featuring a 
quinoid core using density functional theory (DFT), as well as a semi-empirical interacting model Hamiltonian 
describing the $\pi$-system of the molecule which we treat in the GW approximation.  Molecules with a quinoid 
type structure are shown to have two distinct destructive QI features close to the frontier orbital energies. These 
manifest themselves as two dips in the transmission, that remain separated, even when either electron donating 
or withdrawing side groups are added. We find that the position of the dips in the transmission and the frontier 
molecular levels can be chemically controlled by varying the electron donating or withdrawing character of the 
side groups as well as the conjugation length inside the molecule. This feature results in a very high 
thermoelectric power factor $S^2G$ and figure of merit $ZT$, where $S$ is the Seebeck coefficient, making 
quinoid type molecules potential candidates for efficient thermoelectric devices. 
\end{abstract}
\maketitle

\section{Introduction}
Calculations suggest that quantum interference (QI) effects in molecules give rise to a large tunability of the 
electron transport properties of molecular junctions. These effects induce a strong variation of the transmission 
with energy, which is favourable for thermoelectricity, where an electric current or potential difference develops in 
response to a temperature difference across the molecule \cite{pd_2003, bss_2010, klw_2011}. 
Well studied QI molecular units usually involve either cross conjugation and/or meta coupled phenyl units 
\cite{sj_1988, pcg_1997, mwf_2003, sar_2008, sbr_2011}. Molecules such as anthraquinone, have recently been 
shown to exhibit destructive QI effects in a molecular junction \cite{gvm_2012, rcl_2013}. A simple nearest-
neighbour tight-binding model of the $\pi$ system has been used to  rationalise the QI effects. Moreover, 
schemes have been developed to make predictions based on simple graphical rules \cite{mst_2010,mst_2011}. 
The sign of the thermopower has been suggested to provide information of whether the transport is mainly via 
occupied -or unoccupied molecular states \cite{pd_2003}. To obtain a strong thermoelectric response, the 
destructive QI feature needs to be close to the Fermi level. A sensible handle to control the position of QI features 
is offered by tuning the chemistry of the binding groups or changing the electronegativity of substituent side 
groups, which has been shown theoretically as well as experimentally to influence the thermopower by changing 
the molecular frontier levels relative to the Fermi level of the metal electrodes \cite{rjm_2007, bms_2008, 
brg_2012, ega_2013, wcv_2013}. 

Cruciform oligo(phenylene ethynylene)s (OPEs) with a conjugated and extended tetrathiafulvalene (TTF) donor 
moiety have recently been synthesised \cite{wll_2012, pwn_2013} and shown to be redox active as well as having 
interesting spin properties in the Coulomb blockade regime \cite{flvz_2012}. Cruciform type molecules have also 
been synthesised with substituent side groups, such as TTF,  dithiofulvalene (DTF) and atomic oxygen, forming 
cross-conjugated OPEs. Such structures can be referred to as quinoid, since the central core corresponds to a 
quinone, with the substituents replacing the oxygen atoms, see Figure \ref{fig.fig1}a and Figure 
\ref{fig.molecules}. They may exist as zwitterions with mixed substituents, an electron donor on one side and an 
electron accepting group on the other side. This suggests the possibility of electric field induced switching 
between the conjugated (high conductance) and cross-conjugated (low conductance) state. While previous 
theoretical studies have found that the thermoelectric response of molecules may be greatly enhanced by QI 
effects, they involved either rather long molecular wires, radicals or metal complexes without anchoring groups
\cite{sm_2011, bss_2010, klw_2011}.

Here we explore the electron transport properties of cruciform molecules with a quinoid type structure by varying 
the electron donating (ED) and electron withdrawing (EW) character of substituent side groups. We study the low 
bias conductance $G$  as well as the ability of the molecules to convert thermal energy into electric energy by 
applying a temperature difference across the molecular junction. We find, based on density functional theory 
(DFT) calculations, that quinoid structures show two characteristic destructive QI (DQI) features, one near the 
highest occupied molecular orbital (HOMO) level and one near the lowest unoccupied molecular orbital (LUMO) 
level. The DQI features result not only in a high thermopower (Seebeck coefficient) $S$, but more importantly 
from the point of view of technological application, a high power factor (PF), $S^2G$. The PF is related to the 
electrical work that can be extracted from a thermoelectric device and is the quantity that determines the 
dimensionless figure of merit, $ZT=S^2GT/(\kappa_\mathrm{ph} + \kappa_\mathrm{el})$, when the phonon 
thermal conductance $\kappa_\mathrm{ph}$ dominates over the electron thermal conductance $\kappa_
\mathrm{el}$ \cite{m_2013, m_2006a, wcd_2007}. We find that a semi-empirical model for the $\pi$ system, 
treated in the GW approximation \cite{h_1965,tr_2008, srt_2011, st_2012}, gives similar maximal PFs as the DFT 
calculations. We note that a simple nearest-neighbour tight-binding or H{\"u}ckel model does not capture the split 
interference feature. Quinoid type structures typically yield a maximal PF an order of magnitude higher than 
similar molecules showing DQI in the centre of the HOMO-LUMO gap, such as meta coupled benzene or simple 
acyclic cross conjugated molecules.

\section{Method}
We use DFT as implemented in the GPAW and ADF codes to provide an quantum chemical description of the 
charge transport through the molecular junction system \cite{gpaw_2010,vt_2012}.  We also use a more 
approximate density functional tight-binding (DFTB) method \cite{rsp_2007}. The three DFT based methods allow 
us to get an estimate for how sensitive the results are to the particular implementation details. Molecules were 
optimised in the gas phase using the LDA exchange correlation (xc) functional in 
ADF and the PBE \cite{pbe_1996} xc functional in GPAW. Structures used in the DFTB method were relaxed 
using the B3LYP xc functional.
In all calculations the molecules were attached to the FCC hollow site of Au(111) with a Au-S bond length of 
$2.5$ \AA~(1.83~{\AA} above the surface). In GPAW, the scattering region supercell 
was modelled modelled using 3-4 atomic Au layers on both side of the molecule. 
The number of surface layer atoms varies between $4\times4$ and $6\times6$ depending on the
size the molecule and periodic boundary conditions where used in the transverse directions. The 2D Brillouin 
zone was samples using $4\times4$ $k$-points. In ADF the extended molecular region includes $3\times 3$ Au 
atoms in the surface and no periodic boundary conditions are used in the transverse directions. The Au atoms 
were frozen in the bulk lattice structure using the DFT derived lattice constant (PBE: $a=4.18$~\AA, LDA: 
$a=4.08$ \AA).

For calculations based on GW, a semi-empirical model Hamiltonian based on the Pariser-Parr-Pople (PPP) type 
for describing the $\pi$-system is used \cite{p_1953}. We use the Ohno parametrisation \cite{o_1964} with long 
range two-electron interactions with the parameter $U=10.0$~eV describing the Coulomb repulsion and nearest 
neighbour hopping element of $t=-2.5$~eV. On site energies are taken relative to carbon, which we set to $
\varepsilon_c$=0. For the GW calculations we use a wide band approximation for the leads. More details about 
the GW method can be found in Reference \onlinecite{tr_2008, srt_2011, st_2012}.

Transport properties are for all methods calculated using the Landauer B{\"u}ttiker transmission formula 
expressed in terms of Green's functions $\tau(\varepsilon)=\text{Tr}[G^r(\varepsilon)\Gamma^L(\varepsilon) 
G^a(\varepsilon) \Gamma^R(\varepsilon)]$, where $\Gamma^\alpha = i(\Sigma^r_{\alpha} - \Sigma^a_{\alpha})$ 
is given in terms of the lead $\alpha$ self-energy $\Sigma_\alpha$. We calculate the isothermal conductance in 
the zero bias voltage limit as $G=G_0\int (-n_F'(\varepsilon,T)) \tau(\varepsilon) d\varepsilon $, where 
$n_F'(\varepsilon,T)$ is the derivative of the Fermi function with respect to energy. $T$ is the temperature and 
$G_0=2e^2/h$ is the unit of quantum conductance, where $h$ and $e$ are Planck's constant and the electronic 
charge, respectively. We calculate the thermopower in the limit of low temperature drop across the molecule from 
the transmission as $S=(2e/hT) \int \varepsilon (-n_F'(\varepsilon,T)) \tau(\varepsilon) d\varepsilon/ G$. The 
power factor (PF) is calculated as $S^2G$ and we use room temperature $T=300$~K, see Appendix \ref{sec.A2} 
for more details. We note that we do not use the Sommerfeld expansion expression for the 
calculations of the thermopower, since the transmissions functions we consider may have structure on the scale 
of $\mathrm{k_BT}$ \cite{lf_2005}.

\section{Results}
\begin{figure}[h!]
\includegraphics[width=0.95\linewidth]{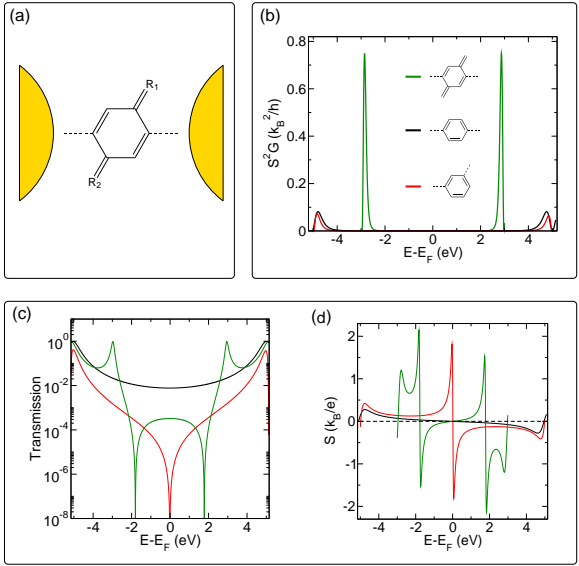}
\caption{\label{fig.fig1}(a) Sketch of a quinoid molecule sandwiched between metal leads. $R_1$ and $R_2$ 
denote the substituent side groups. (b) Powerfactor $S^2G$ for a quinoid structure and benzene connected 
in ortho and meta position. The insets shows the chemical structures of the molecules.  The results are obtained 
using a PPP model  with GW. (c) and (d) show the transmission and thermopower as a function of energy.}
\end{figure}

We begin by considering a simple model for a quinoid structure, 
which captures the essential physical mechanism responsible for interesting thermoelectric properties of quinoids. 
The schematic structure of a transport junction with a quinoid molecule sandwiched between two leads i shown 
in Figure \ref{fig.fig1}a. The connection of the central unit to the leads are indicated by the dashed lines. $R_1$ 
and $R_2$ denote substituent side groups. In Figure \ref{fig.fig1}b we show the PF ($S^2G$) calculated with GW 
for a PPP model Hamiltonian where $R_1$ and $R_2$ are taken as CH$_2$ groups, 
as indicated by the inset. The model includes a single $p_z$ orbital per carbon atom all with equal on site
energies. For comparison we also show the PF obtained for benzene coupled in para and meta position.  The 
maximal PF, within the HOMO-LUMO gap, for the quinoid structure is seen to reach a value that is an order of 
magnitude higher than both meta and para coupled benzene. In fact the PF for the quinoid is comparable 
to the maximal PF that can be obtained from a Lorentzian transmission line shape, $\sim 0.9~\mathrm{k_B^2}/h$,
by using a width of about $1.1~\mathrm{k_B}T$, see Figure \ref{fig.lorentzian_model}a in Appendix \ref{sec.A2}.
%The broadening for the simple model is $\sim10\mathrm{k_B}T$.

The PFs can be analysed in terms of the transmission and thermopower as a function of
energy, shown in Figure \ref{fig.fig1}c and \ref{fig.fig1}d, respectively. The transmissions for both the quinoid and 
the meta coupled benzene ring is seen to be highly suppressed, as compared with para coupled benzene, for an 
extended energy region around the Fermi level. 
This is a result of DQI, as we will discuss in more detail below. DQI also introduces a strong variation of the 
transmission with energy which in turn yields a high thermopower. This can be seen by considering the linear 
response Sommerfeld expansion expression for the thermopower $S(\varepsilon)\propto \partial_\varepsilon \log 
[T(\varepsilon)]$ \cite{lf_2005}.
The meta-coupled benzene ring has a single dip in the transmission at the centre of the HOMO-LUMO gap, while 
the quinoid structure shows two transmission dips -- one dip is close to the HOMO energy while the other dip is 
close to the LUMO energy.  While the thermopower for meta-coupled benzene reaches high values close to the 
Fermi level, the PF ends up being low because the transmission, and thus $G$, is very low here. 
The quinoid structure on the other hand, has the DQI feature near a transmission resonance giving both a high 
thermopower and transmission within the same energy region. This combination is responsible for the high PF for 
the quinoid structure. The para-coupled benzene shows a relatively low thermopower, but relatively high 
transmission 
and thus ends up having a PF very similar to the meta-coupled benzene.
Finally we note that the quinoid structure has a smaller HOMO-LUMO gap, which is related to the longer 
conjugation length as
compared with benzene, \emph{i.e.} it is a confinement effect. 
The maximal PF value is obtained at an energy located between the HOMO resonance and the dip, which is 
about $0.2$~eV away form the HOMO resonance for all three molecules.

We now analyse how the thermoelectric properties can be improved by adding side groups to the quinoid 
backbone. The aim is to move the DQI feature close to a transmission resonance (HOMO or LUMO) which leads 
to a peak in the PF, and to shift that peak towards the metal's Fermi energy. 

\begin{figure}[h!]
\includegraphics[width=0.95\linewidth]{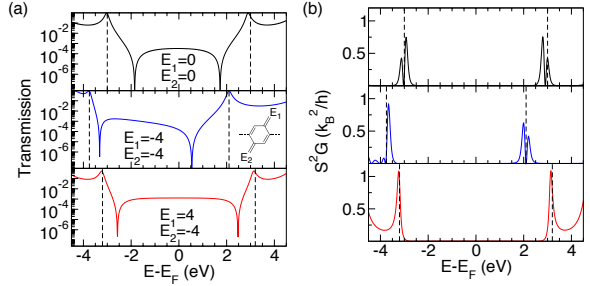}
\caption{
\label{fig.model_trans_s2g}
(a) Transmissions for the quinoid type structure with on-site energies ($E_1$, $E_2$) of the ($R_1$, $R_2$) side 
groups of (0.0, 0.0) top, (-4.0,-4.0) middle and (-4.0,+4.0) bottom in units of eV. The vertical dashed lines indicates 
the HOMO and LUMO levels. (b) Same as in  (a) but for the power factor, $S^2G$.}
\end{figure}

To probe how the two DQI features depend on the choice of the side groups, we show in Figure 
\ref{fig.model_trans_s2g}a and \ref{fig.model_trans_s2g}b how varying the on-site energies of the side groups in 
the model Hamiltonian affects the transmission and PF, respectively. 
The side group site energies are 0.0~eV for both $R_1$ and $R_2$ in the top panel, -4.0~eV for both $R_1$ and 
$R_2$ in the middle panel and -4.0 and 4.0~eV for $R_1$ and $R_2$ in the lower panel. 
The negative/positive site energies are meant to mimic the effect of EW/ED side groups. We note that ED 
groups (EDG) and EW groups (EWG) will in general also tend to shift the on-site energies of the ring through 
inductive and resonance effects by perturbing the electron density within the ring in the $\sigma$ and $\pi$-
system, respectively. The perturbation of the electron density through the $\sigma$-system is caused by 
polarisation and is not included in the PPP model. 
The change in electron density will in turn affect the electrostatic potential and can be seen as a way of 
chemically gating the molecular backbone. 

The position of the transmission dips is seen to shift with the side group energies but the two split QI features are 
quite robust and stay within the HOMO-LUMO gap. It can be seen that a transmission dip can be moved
closer to the HOMO resonance by lowering the side group energies, which leads to an increase in the 
corresponding PF. This can be rationalised from the approximate expression $S\propto\partial_\varepsilon 
\log[T(\epsilon)]$ for the thermopower,
that is, the slope of the transmission as function of energy on a logarithmic scale. 
The slope of the transmission is seen to increase in Figure \ref{fig.model_trans_s2g}a when on-site energies are
varied such that the transmission anti-resonance moves closer to the molecular frontier energy levels, \emph{i.e.}
the transmission has to change from a value of 1 at resonance to 0 at the antiresonance over a smaller energy 
region. The calculated maximal PF within the HOMO-LUMO gap are high in all cases, and reach a value of 
$\sim1.1~\mathrm{k_B}^2/h$ (=$316$~fW/K$^2$). A similar value was
obtained by optimising a general two level model in Reference \onlinecite{klw_2011}. 
We therefore expect that the EDGs and EWGs can not only be used to tune the position of the interference dips, 
but also to enhance the PF.
To experimentally obtain high PFs, molecules are needed where either the HOMO or the LUMO level are close to 
the Fermi level. This can be chemically controlled by the binding (anchoring) groups \cite{clt_2006, psm_2009, 
ggw_2013}, or by mechanical tuning \cite{pvd_2013}, as well as by the EW/ED  side groups. Before exploring a 
more detailed  quantum chemical description of the molecules shown in Figure \ref{fig.molecules}, we analyse the 
two split interference features in more detail for the model of the quinoid core structure.

The appearance of the QI features can be analysed in a number of ways ranging from non-spanning nodes to 
interfering pathways through Feynman path integrals and the phases of molecular orbitals \cite{lee_1999, 
th_2014, sss_2014, yts_2008, sar_2008}. Here we rationalise the QI interference features in terms of molecular 
orbitals and corresponding molecular energies. This has the advantage that in the weak molecule lead coupling 
limit this analysis can be generalised for the correlated electron case by considering so-called Dyson orbitals and 
energies \cite{psp_2014}.

We argue that the central quinoid core structure is responsible for the two transmission nodes.
This can be seen either based on graphical rules developed to predict QI in H{\"u}ckel models \cite{mst_2010} or 
by systematically reducing the Kohn-Sham Hamiltonian atom by atom as will be discussed below. We note that 
while a nearest neighbour H{\"u}ckel model predicts degenerate transmission anti-resonances for equal site 
energies \cite{mst_2011}, both the GW  and the DFT calculations give two well separated anti-resonances for 
quinoid structures. A similar effect was observed and formulated in terms of long range hopping elements 
for acyclic cross-conjugated molecules in Reference \onlinecite{sbr_2011}.

For simplicity we proceed with an effective non-interacting model where we use the H\"{u}ckel molecular orbitals 
but quasi-particle energies taken from the GW calculations. The use of quasi-particle energies instead of 
H\"{u}ckel molecular yields split interference features \cite{psp_2014}. 
The two central DQI features of the quinoid structure, see Figure \ref{fig.fig1}a,c, can be accounted for by 
considering the four molecular orbitals closest to the Fermi level (HOMO-1, HOMO, LUMO, LUMO+1). This may 
be seen by first considering the
condition for the transmission going to zero, $T(\varepsilon)\propto|G_{lr}(\varepsilon)|^2=0$. 
Here the Green's function $G_{lr}$ describes the probability amplitude for an electron or hole propagating through 
the molecule to a site $l$ from a site $r$ on the molecule, which are connected to the left and right lead, 
respectively. 
Then, by expressing $G_{lr}$ in terms of the molecular (uncoupled) Green's function as $G_{lr}(\varepsilon)=G^
\text{mol}_{lr}(\varepsilon)/D(\varepsilon)$ \cite{ccs_1971}, 
we see that a necessary condition for a transmission zero 
is that $G^\text{mol}_{lr}$ is zero for some energy. The denominator in the expression for $G_{lr}$ is given by  
$D=(1-[\Sigma_{L}]_{ll} G^\text{mol}_{ll})(1-[\Sigma_{R}]_{rr}G^\text{mol}_{rr})$. 
In the spectral representation $G^\text{mol}_{lr}$ may be expressed in terms of molecular orbitals 
$\{\psi_n\}$ and energies $\{\varepsilon_n\}$ as
\begin{equation}\label{eq.glr}
G^\text{mol}_{lr}(\varepsilon) = \sum_n \frac{\langle l| \psi_n \rangle\langle \psi_n | r\rangle}{\varepsilon-
\varepsilon_n+i\eta} ,
\end{equation} 
where $\eta$ is a positive infinitesimal. 
The elements $\langle l |  \psi_n \rangle$ and $\langle r | \psi_n \rangle$ give the amplitude  of the
$n$'th molecular orbital on site $l$ and $r$, respectively. 

\begin{figure}
\includegraphics[width=0.95\linewidth]{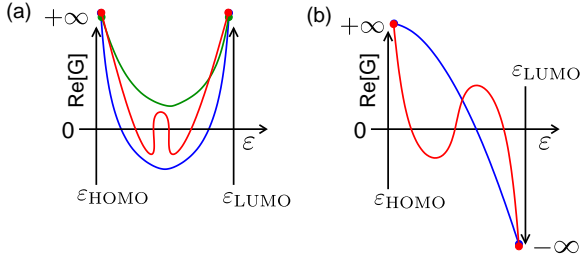}
  \caption{\label{fig.tzeros} (a) Sketch of a few representative cases of Re[$G^\text{mol}_{lr}$] from HOMO and 
LUMO orbitals having different relative sign on sites connected to leads. (b) same as (a) but for HOMO and 
LUMO orbitals with same sign on sites connected to leads. }
\end{figure}

From Eq. \eqref{eq.glr} we infer that the relative sign of the HOMO- and LUMO amplitudes on the 
sites $l$ and $r$ 
determine the parity of the number of transmission zeros within the HOMO-LUMO gap. 
This is illustrated in Figure \ref{fig.tzeros}a and \ref{fig.tzeros}b, where we for a few representative cases sketch 
the real part for of $G^\text{mol}_{lr}$ within the HOMO-LUMO gap with an even and odd number of zero 
crossings, respectively. The contribution to Re[$G_{lr}^\text{mol}$] from the HOMO and LUMO diverges when 
approached from within the HOMO-LUMO gap towards either $+\infty$ or $-\infty$, with the sign determined by 
the residue (=$\langle l |\psi_n \rangle \langle \psi_n |r\rangle \rangle$). 
All other orbitals only contribute with a finite value to Re[$G_{lr}^\text{mol}(\varepsilon)$] within the HOMO-LUMO 
gap. 
This means that the function Re[$G_{lr}^\text{mol}(\varepsilon)$] can in general 
be drawn as a continuous line connecting the divergence at $\varepsilon_\text{HOMO}$ and $\varepsilon_
\text{LUMO}$, with the detailed shape determined by all HOMOs and LUMOs. 
If the residues of the HOMO and LUMO have different signs we are 
in the situation corresponding to Figure \ref{fig.tzeros}a, where we see that any continuous line is forced to cross 
zero and even number $(0, 2, 4,...)$ of times. With four orbitals, as in our model, only 0 or 2 zero crossings are 
possible. 

If the the HOMO and LUMO residues have the same sign, we are in the situation shown in Figure \ref{fig.tzeros}b, 
where any continuous line is forced to cross zero an odd $(1, 3, 5,...)$ number of times. From this a quick 
assessment and classification of interference features can be made in terms of the HOMO and LUMO orbitals 
only: Constructive interference between HOMO and LUMO results in an even number of transmission zeros while 
for destructive interference the number of zeros is odd.

\begin{figure}[h!]
 \includegraphics[width=0.95\linewidth]{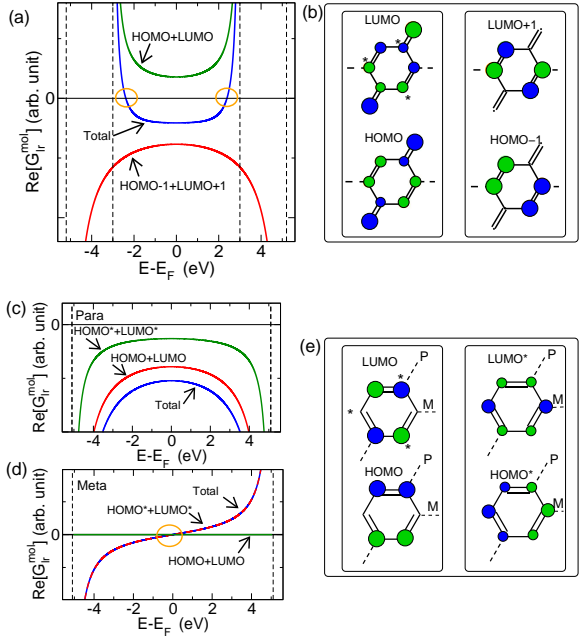}
 \caption{\label{fig.poles} (a) Contributions to the real part of the molecular Green's function $G^\text{mol}_{lr}$ for 
the quinoid structure; HOMO and LUMO (blue curve) pair and the HOMO-1 and LUMO+1 pair (red curve). The 
vertical dashed lines indicate the positions of the poles, arising from the molecular energy levels. (b) The 
molecular orbitals are represented with the weight on the different sites by the size of the circles and the sign by 
the colour. (c) and (d), same as (a) but para and meta connected benzene, respectively. (e) same as (b) but for 
benzene para.}
\end{figure}

\begin{figure}[h!]
\includegraphics[width=0.4\linewidth]{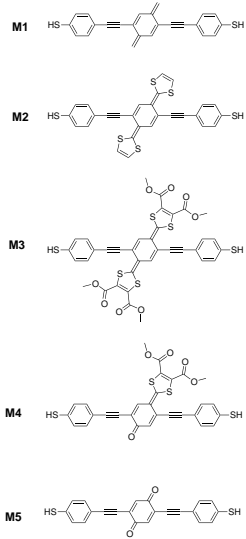}
\caption{\label{fig.molecules} Molecules M1-M5 with a quinoid type structure consisting of a 
central cross-conjugated  phenyl ring. M1 contains CH$_2$ side groups,  M2 contains a TTF electron donating 
unit, M3 contains a TTF unit with esters, M4 is a possible zwitter-ion with an electron donating DTF unit with ester 
groups on one side, and an oxygen atom on the other side. Finally M5 contains two electron withdrawing oxygen 
atoms.}
\end{figure}

To simplify the analysis even further we now turn to the Coulson-Rushbrooke (CR) pairing theorem \cite{cr_1940} 
for alternating hydrocarbons, \emph{i.e.} molecules which can be viewed as
bipartite. The pairing theorem relies on a particle-hole symmetry and states that the molecular energies come in 
pairs as $\varepsilon_{\text{HOMO}-n}=\varepsilon_{\text{LUMO}+n}$ for $n=0, 1, 2,...$ . 
The molecular orbitals of such a pair are identical except for a sign change on one of the sub-lattices, \emph{i.e.} 
every other atom, which means that the sign of the residues of the orbitals in a CR pair can be predicted without 
any calculations. In short, if leads are connected to the same sub-lattice of a molecule the orbitals in a CR pair, 
they interfere destructively, giving an odd number transmission 
zeros. 
For leads connected at sites belonging to different sub-lattices, the two orbitals in a CR pair
interfere constructively, giving an even number of transmission zeros.  
In Figure \ref{fig.poles}b, we sketch the four important quinoid frontier molecular orbitals, with the weight on a site 
represented by the size of the circle and the sign by the color. The dashed lines indicate the sites $l$ and $r$, for 
which leads are connected. 
The starred sites constitute one sub lattice of the molecule. We see that the paired orbitals, namely the HOMO 
and LUMO and also the HOMO-1 and LUMO+1, indeed follow the behaviour predicted by the CR theorem.
Due to the close relation between orbitals in a CR pair, we find it advantageous to first sum up the contribution 
from a CR pair in Eq. \eqref{eq.glr} and subsequently sum up all pairs.

In order to enable destructive interference in the case where CR-paired orbitals interfere constructively, two or 
more sets of such paired orbitals are needed, \emph{i.e.} a minimum of four molecular orbitals.
We show the paired HOMO-LUMO (blue line) and the paired HOMO-1 and LUMO+1 (red line) contributions to 
$\text{Re}[G^\text{mol}_{lr}]$ for the quinoid in Figure \ref{fig.poles}a. 
Here the vertical dashed lines indicate the positions of the quasi-particle levels. It can be seen that the HOMO-
LUMO pair adds up constructively due to the different signs arising from the product of the
molecular orbital weights on the $l$ and $r$ sites -- a consequence of the leads being connected to different sub-
lattices. In the same way the HOMO-1 - LUMO+1 pair interferes constructively. 
The destructive interference observed in the total transmission 
arise when we consider the sum of the pair contributions (green line),
which is seen to cross zero twice in accordance with the general classification based on the amplitudes of the 
HOMO and LUMO orbitals on the sites connected to leads.
The transmission nodes are a consequence of the HOMO and LUMO pair having a smaller weight on the $l$ and
$r$ sites than the HOMO-1 and LUMO+1 pair. If we imagine increasing the HOMO and LUMO orbital weight on 
the lead sites, then at some point the transmission nodes will disappear, although the transmission will still be 
suppressed. The DQI is for the quinoid therefore not only a result of the phases of the molecular orbitals. 
Similarly the relative position of the HOMO- and HOMO-1 level (and the LUMO-
 and LUMO+1 level) also influence the appearance of transmission zeros.

It is interesting to compare the quinoid to benzene in para position, where the orbitals making up a CR pair 
interfere constructively just as for the quinoid, see Figure \ref{fig.poles}c and \ref{fig.poles}e. However, in contrast 
to the quinoid, adding up the two pairs now leads to constructive interference. This is a
result of a different phase relations between the two CR pairs for benzene. 
For meta coupled benzene, the CR pairing theorem ensures that the two orbitals making up a CR pair always 
interfere destructively, resulting in a zero crossing in $\text{Re}[G^\text{mol}_{lr}(\varepsilon)]$ at the Fermi level, 
see Figure \ref{fig.poles}d.

As already mentioned, a nearest neighbour H{\"u}ckel or tight-binding model, predicts a doubly degenerate node, 
and does not capture the characteristic split interference feature of the quinoid structure \cite{mst_2011, 
psp_2014}. In terms of the analysis presented here, which was based on Eq. \eqref{eq.glr}, the doubly 
degenerate transmission node can be interpreted graphically as in Figure \ref{fig.poles}a. 
$\text{Re}[G^\text{mol}_{lr}]$ (blue line) may be shifted up along the $y$-axis by a 
change in the  molecular energy levels (or a change in orbital weight on $l$ and $r$). At some point, the two zero 
crossings coincide at the Fermi level. This is exactly what happens if one uses the tight-binding molecular orbitals 
and energies to construct $G^\text{mol}_{lr}$. The splitting can be obtained by adding a scissors operator to the H
\"{u}ckel  Hamiltonian to change the HOMO-LUMO gap. The scissors operator will in general have long range 
hopping matrix elements, which has been shown to split interference features in some cases \cite{sbr_2011}.

From the four frontier molecular orbitals in Figure \ref{fig.poles}b, we can estimate the effect of changing the on-
site energies on the side groups. The HOMO-1 and LUMO+1 have (almost) zero weight on the side group sites 
and a perturbation of the on-site energies, $\hat{V}_{sg}$, will therefore not change the HOMO-1 and LUMO+1 
orbitals and energies to lowest order since  $\hat{V}_{sg} | \alpha \rangle \approx 0$ for $\alpha \in\{$HOMO-1, 
LUMO+1$\} $. Thus, the HOMO- and LUMO level shift a lot, but the HOMO-1 and LUMO+1 do not.
This can be used to give a graphical interpretation in Figure \ref{fig.model_trans_s2g}a. 
A change of $\Delta \varepsilon$ of the side group on-site energies will result in a shift of the HOMO- and LUMO 
levels of $\Delta \varepsilon$, while the HOMO-1- and LUMO+1 levels stay fixed. 
This will result in a shift of the position of the two dips in the same direction as the HOMO and LUMO level.

To explore the robustness as well as the tunability of the DQI features with respect to the EW/ED character of the 
substituent side groups we have performed quantum chemical calculations based on DFT 
for the molecules M1-M5 shown in Figure \ref{fig.molecules}.

\begin{figure}[h!]
 \includegraphics[width=0.95\linewidth]{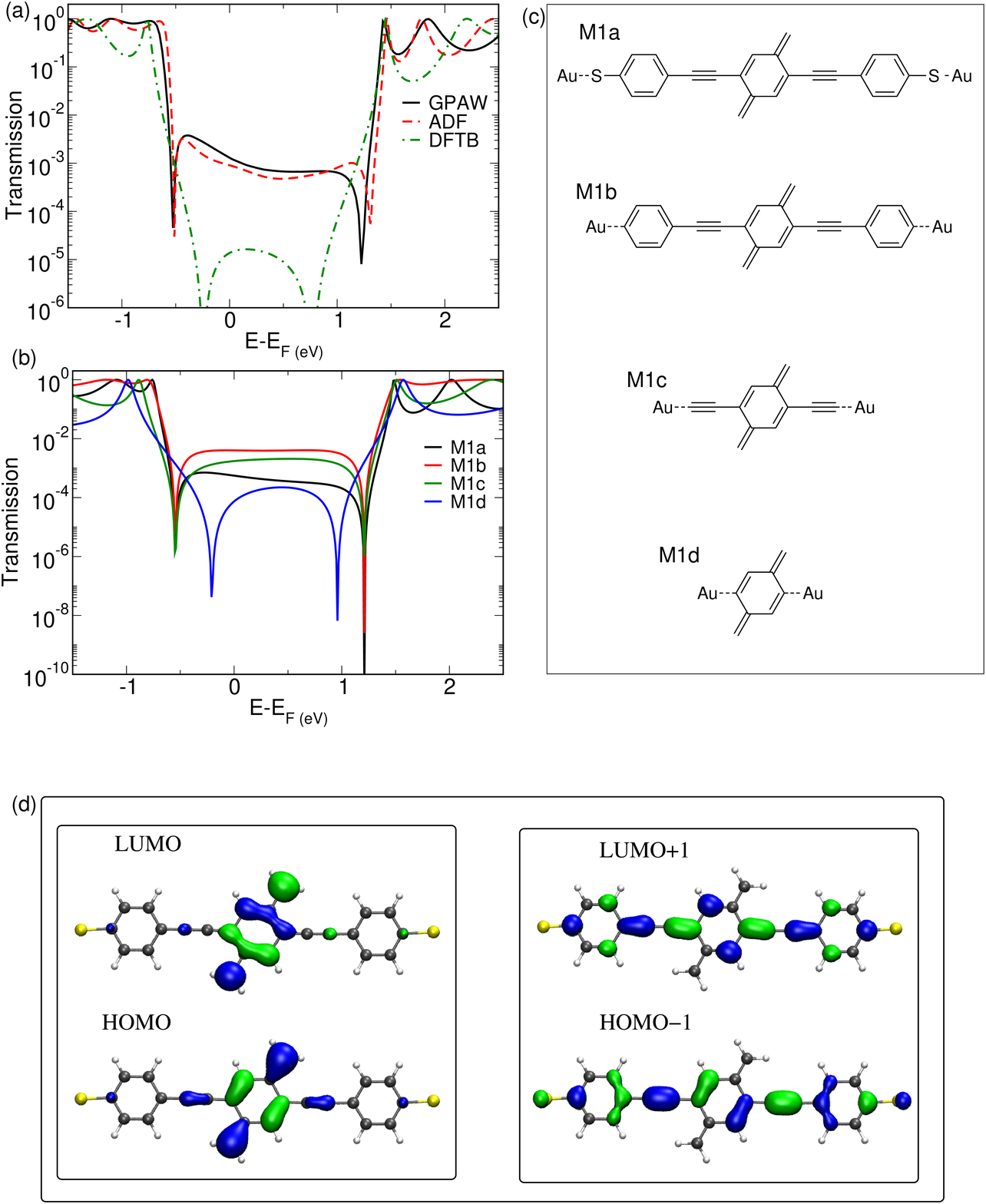}
  \caption{\label{fig.trans_m1_compare} (a) Transmission as a function of energy for molecule M1 calculated with 
GPAW, ADF and DFTB. (b) Transmissions within the $p_z$ subspace but for truncated versions of M1 as denoted 
in (c). Wide band leads are attached at the $p_z$ orbitals connected to "Au" with a dashed line in (c). (d) Frontier 
molecular orbitals for M1.}
\end{figure}

Before discussing molecules M1-M5 in more detail, we give a comparison between the transmission calculated 
with the different DFT methods in Figure \ref{fig.trans_m1_compare}a for M1. The two split DQI features 
discussed for the simple quinoid model above are clearly visible and we emphasise the very good agreement 
between the GPAW and ADF calculations despite small variations in geometry and the use of different xc 
functionals. We note that the agreement between the methods,
especially the position of the Fermi level, is less good for M2-M5. However, the side group induced changes and 
trends are not affected and we therefore proceed discussing the results obtained with GPAW. 

We stress that we have used flat Au(111) surfaces for simplicity and not investigated the effect of using different lead structures, such as a small pyramids or add atoms on a flat surface. 
While the use of different lead structures may certainly change the conductance and the 
thermoelectric properties we do not think it will change the trends and conclusions as these are based on a property of the molecule itself.

To make a more direct connection to the minimal $p_z$ model for the quinoid core structure, we have performed 
calculations starting from the molecule M1 and then systematically removed atoms from its arms.
The transmissions in Figure \ref{fig.trans_m1_compare}b are obtained by only considering the $p_z$ subspace of 
the Kohn-Sham Hamiltonian of the full calculation for successive truncations of the arms of the molecule M1, 
while keeping the central cross-conjugated quinoid unit unchanged. The different transmissions correspond to the 
molecules denoted M1a-M1d shown in Figure \ref{fig.trans_m1_compare}c. Wide band leads are attached to the 
atoms connected with a dashed line. The position of the transmission nodes are seen to be quite insensitive to 
the arms of the molecule attached to the central quinoid core structure. Only when leads are attached directly to 
the quinoid core structure do we see a non-negligible change. We stress that in all cases the qualitative shape of 
the transmission function remains intact and in good agreement with the GW results in 
Figure \ref{fig.model_trans_s2g}a.

\begin{figure}%[h!]
 \includegraphics[width=0.95\linewidth]{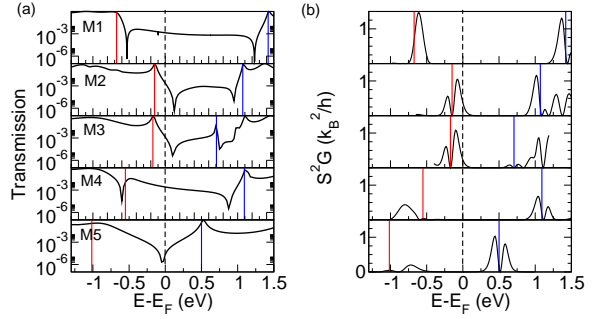}
  \caption{\label{fig.dft_trans_s2g} (a) Transmissions for molecule M1-M5 calculated with DFT (GPAW). (b) Power 
factor for M1-M5. The vertical colored lines indicate the position of the frontier energy levels of the molecule 
(HOMO: red, LUMO: blue).}
\end{figure}

In Figure \ref{fig.dft_trans_s2g}a and \ref{fig.dft_trans_s2g}b we show the transmission  and PF calculated using 
GPAW (see Appendix \ref{sec.A1} for the corresponding ADF and DFTB transmission results) for the molecules 
M1-M5. 
The vertical lines indicate the position of the molecular frontier energy levels obtained by diagonalising the 
molecular subspace (including sulfur). 

For all molecules, except M5, we see clear signs of the two split DQI features. Interestingly, M5 only shows a 
single transmission anti-resonance, which according to the discussion above based on Eq. \eqref{eq.glr} implies 
that the HOMO and LUMO orbital in this case interfere destructively.
Taking molecule M1 as the baseline, we observe that the HOMO energy level is closer to the Fermi level than the
LUMO energy level. This is typical for molecules bonded to Au via thiols \cite{psm_2009}. The similarity of
the shape of the transmission with the results for the simple quinoid model (see Figure \ref{fig.model_trans_s2g}
a) is striking. 
We see two DQI features, one located near the HOMO and the other near the LUMO level. The PF for M1 is 
shown in the upper most panel in Figure \ref{fig.dft_trans_s2g}b and has a maximal value of $\sim1.5$~$
\mathrm{k_B}^2/h$ within the HOMO-LUMO gap. This is 
larger than, but comparable to our GW result for the quinoid model and about 60\% higher than the maximal value 
of a single level model with a Lorentzian line shape, see Appendix \ref{sec.A2}. Interestingly, the maximal PF 
for M1 is even larger than the optimized value obtained for a general two-level non-interacting model
\cite{klw_2011}. The high PF obtained 0.75 eV below the Fermi level is a result of the transmission anti-resonance
being close to a resonance such that both the thermopower and conductance are high at the same energy, as 
already discussed for the simple quinoid model. 
In Figure \ref{fig.trans_m1_compare}d we show the frontier molecular orbitals for 
M1 which are responsible for the two split transmission nodes. The orbitals can be compared with the simple 
model in Figure \ref{fig.poles}b and are seen to have the same essential features: The HOMO and LUMO orbitals 
have smaller weight at the connecting sites than the HOMO-1 and LUMO+1 orbitals, and the orbitals in a CR pair 
interfere constructively. 
Also the symmetry of the orbitals with respect to the transport direction are the same.
The essential mechanism behind the two DQI features in the DFT
calculations is therefore captured by the simple PPP semi-empirical $\pi$-system 
model for the quinoid core structure. The PF for M1 at the Fermi level is rather low, suggesting that the band 
alignment of the molecular levels with the Fermi level of gold is not ideal in terms of possible thermoelectric 
technological applications.

Molecule M2 contains two electron-rich five membered rings containing sulfur (DTFs) that have an ED character.
These units are therefore expected to push the molecular levels up in energy as compared with M1 without the 
ED rings. Indeed, we see that the HOMO -and LUMO levels are moved up in energy with the HOMO level ending 
up very close to the Fermi level. We note that with the GPAW code, the HOMO level ends up just below the Fermi 
level, while it ends up above the Fermi level with the  ADF code. We have verified that this is not a basis set 
issue, by comparing to real space grid calculations performed with GPAW. Based on this we speculate that this 
difference is due to use of a cluster model instead of a periodic structure for the leads in ADF.
The HOMO-LUMO gap of M2 is considerably reduced as compared with M1. This is an effect of the increased 
conjugation length of the TTF side groups in M2 compared with the CH$_2$ side groups in M1. However, the two 
split DQI features are still intact and within the HOMO-LUMO gap. 
While the maximal PF is similar to M1, the PF at the Fermi level is now relatively high, with a value of $0.4$ k$_
\text{B}^2/e$ ($\sim 100~\mathrm{fW/K}^2$).

Molecule M3 is similar to M2 but with esters added to the DTF substituents which may have a beneficial effect on 
the solubility of the molecule.
The addition of esters alters the transmission spectra slightly near the Fermi level as compared to M2.
However, there are now unoccupied transmission resonances at higher energies, such that the high-energy anti-
resonance now falls outside the HOMO-LUMO gap. These transmission resonances do not reach a value of 1 
and originate from two orbitals, LUMO and LUMO+1, asymmetrically coupled to leads and with most of the weight 
localised on the DTF+ester side group. 
The PF at the Fermi level is again rather high and comparable to what we found for the M2 molecule.

The M4 molecule is an interesting molecule and a possible zwitterion. It contains an EDG (DTF) as one side 
group and an EWG (atomic oxygen) as the second side group. This gives the molecule a large dipole moment. 
The transmission as a function of energy shows the two DQI transmission features. 
The HOMO is coupled rather asymmetrically which results in a transmission resonance with a 
value below 1. We see that the opposite character of the two side groups tends to cancel the overall shift of the 
levels, in agreement with the PPP model studied above, see Figure \ref{fig.model_trans_s2g}a.

Molecule M5 has a real quinone core. A similar molecule, namely the anthraquinone was recently shown to 
exhibit DQI features in the dI/dV \cite{gvm_2012, rcl_2013}. 
The two electronegative oxygen atoms are EW which is reflected by the molecular levels being drawn down in 
energy. The LUMO is now closer to $E_F$ than the HOMO. There is clearly only one transmission dip within the 
HOMO-LUMO gap. 
\begin{figure}%[h!]
 \includegraphics[width=0.5\linewidth]{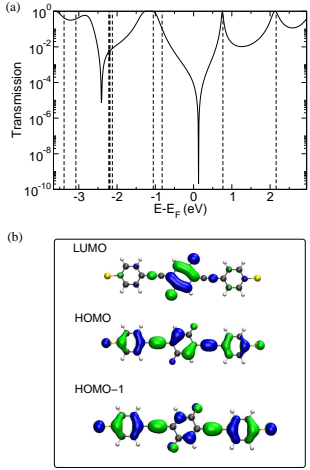}
  \caption{\label{fig.M5_trans_pz} (a) $\pi$-system contribution to the transmission for M5. The vertical dashed 
lines indicates the positions of the molecular energy levels. (b) Frontier molecular orbitals for M5.}
\end{figure}
In Figure \ref{fig.M5_trans_pz}a we show the $\pi$ system contribution to the transmission for M5, from which it is 
clear that a second interference feature is present at $-2.5$~eV below the Fermi level. However, since this is not 
within the HOMO-LUMO gap it is not relevant. By inspecting the frontier molecular orbitals in Figure 
\ref{fig.M5_trans_pz}b we see that that while the LUMO is the same as for M1, the HOMO is not. The symmetry 
of the HOMO and LUMO orbitals now implies destructive interference within the HOMO-LUMO gap. From the 
analysis based on Eq. \eqref{eq.glr},
we expect an odd number of transmission zeros within the HOMO-LUMO gap in this case. This is consistent with 
the observation of one transmission node. The transmission node is positioned close to the Fermi level which 
induces a strong variation of the transmission function and thus a large thermopower. However, the transmission 
is suppressed, and thus the resulting PF at the Fermi level is rather low.
\begin{figure}%[h!]
 \includegraphics[width=0.95\linewidth]{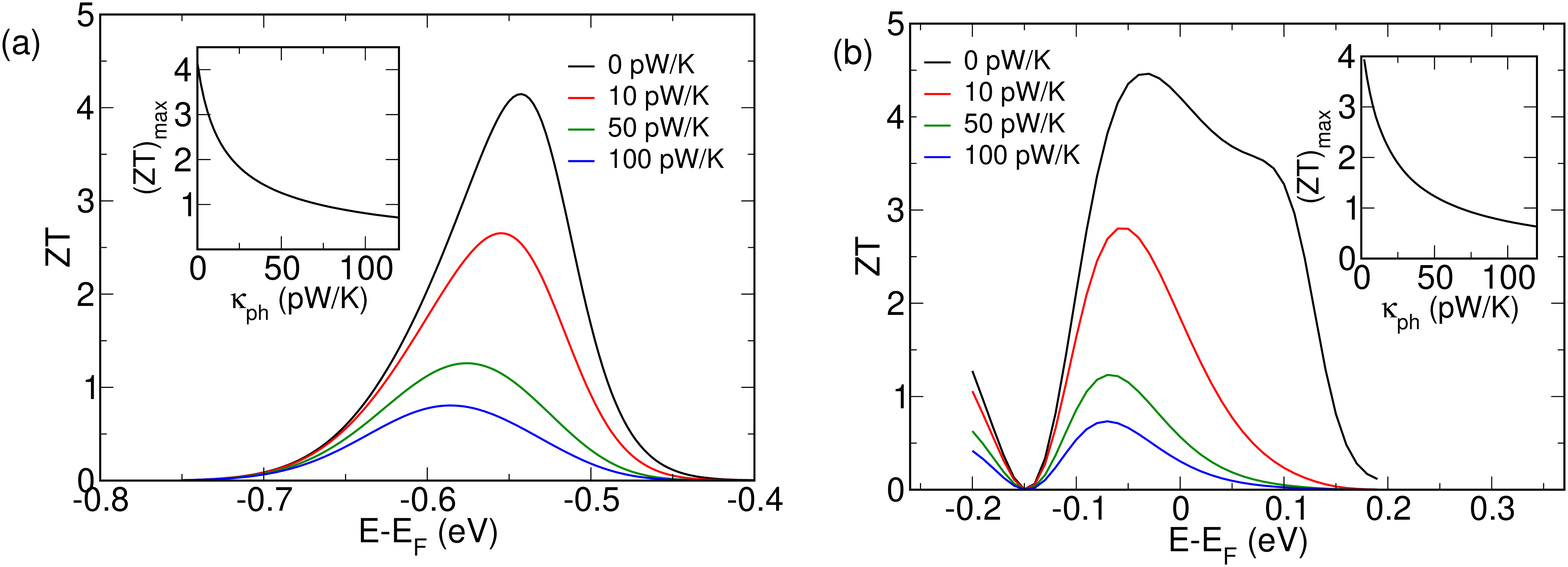}
  \caption{\label{fig.zt} Figure of merit for M1 and M2 is shown in (a) and (b), respectively. 
  The different lines are obtained by using different values for the phonon
  thermal conductance $\kappa_{ph}$, as indicated in the legends.
  The inset shows the maximal value of ZT within the HOMO-LUMO gap as function of 
  the phonon thermal conductance.}
\end{figure}

To investigate the general thermoelectric performance of the molecules we show in Figure \ref{fig.zt}a and 
\ref{fig.zt}b the dimensionless figure of merit ZT as a function of energy for molecule M1 and M2, respectively. To 
probe the dependence on the phonon thermal conductance, $\kappa_{ph}$, we show the results for four different 
values ranging from 0 to 100 pW/K. Thermal conductance values in the range 10-100 pW/K have recently been 
calculated for similar OPE3 molecules \cite{m_2013}.
Measurements on $n$-Alkanes suggest a similar range of values. We see that maximal ZT, denoted $
\mathrm{(ZT)_{max}}$, is almost identical for the two molecules. This is especially clear when comparing the 
insets which show $\mathrm{(ZT)_{max}}$ as a function of $\kappa_{ph}$. For comparison the maximal ZT which 
can be obtained using  a thermal conductance of $\kappa_{ph}=10$~pW/K for a single level model with a 
Lorentzian line shape is about $2.4$, see Appendix \ref{sec.A2}. 
However, in this case a rather narrow resonance is 
needed with a with of about $0.3\mathrm{k_B}T$. The dependence of ZT on energy, suggests that molecular 
levels need to be aligned with the Fermi level with a precision of about $\sim 0.1$ eV. For molecule M2 the 
alignment predicted by DFT (GPAW) is relatively good giving high values for ZT at the Fermi level. 
\begin{table}
  \caption{ZT and PF for the molecules M1-M5 obtained using DFT-PBE. ZT is for $\kappa_{ph}=0, 10$~pW/K. PF 
is in units of $\mathrm{k_B}^2/h$ ($\sim 288 fW/K^2$).}
  \label{tbl.table1}
  \begin{tabular}{llll}
    \hline
    Molecule  & $\mathrm{(ZT)_{max}}$ & $\mathrm{(ZT)_{E_F}}$ & $\mathrm{(PF)_{max}}$ \\
    \hline
    M1   & 4.1, 2.7 & 0.014, 0.001  & 1.5   \\
    M2   & 4.5, 2.8 & 4.2, 1.8 & 1.2  \\
    M3   & 4.1, 2.7 & 3.5, 1.3 & 1.1   \\
    M4   & 5.5, 2.5 & 0.032, 0.004 & 0.7   \\
    M5   & 1.8, 1.5 & 0.706, 0.005 & 1.0  \\
    \hline
  \end{tabular}
\end{table}
We have collected   $\mathrm{(ZT)_{max}}$ 
and ZT evaluated at the Fermi level, $\mathrm{(ZT)_{E_F}}$, and $\mathrm(PF)_\text{max}$ in Table 
\ref{tbl.table1}. 
Here we clearly see that all the quinoids show promising thermoelectric properties when the maximal values are 
considered. In terms of the level alignment predicted by DFT molecules M2 and M3 look promising both in terms 
of power production and efficiency. We note that the level prediction by DFT may not be very accurate, however, 
we believe that the trend in the level position induced by the substituent side group is a robust feature.
Also, we have only considered one type of  binding structure between the sulfur anchoring group and the gold surface. Both the level positions and the broadening of levels may be sensitive to the details of the gold-sulfur interface and this will
in turn affect the calculated thermoelectric properties.
However, if it is mainly the broadening of the levels that are affected by the binding geometry, then we do not expect large changes in the calculated properties. This is because the destructive QI forces the transmission to change from 1 to 0 between 
a frontier orbital resonance and a nearby anti-resonance irrespectively of the broadening. 

\section{Conclusions}
We have shown, based on quantum chemical calculations using DFT as well as GW for an interacting 
semi-empirical Hamiltonian, that molecules with a quinoid topology may show very high power factors and ZT 
values, which suggests a high power generation per molecule and good efficiency, respectively.

The good thermoelectric properties were found to originate from a particular DQI feature of quinoid type 
structures, namely two split interference features within the HOMO-LUMO gap, resulting in a transmission anti-
resonance lying  close to a molecular resonance. This feature was shown to involve the four nearest frontier 
molecular orbitals. The split interference feature was found to be rather robust and the position of the resonances 
and anti-resonances was shown to be highly tuneable by the ED/EW nature of side group substituents.
For the five molecules studied, only the real quinone core breaks the two split interference feature, only having a 
single transmission node in the HOMO-LUMO gap.

DFT calculations showed that the maximal power factor as well as ZT obtainable within the HOMO-LUMO gap
was only weakly dependent on the chemical nature of the side group. However, by varying the ED/EW character 
of the side group the power factor and ZT values evaluated at the Fermi level can be tuned. 
 Three different DFT based methods predicted the same trends, however, the exact position of the mid HOMO-
LUMO gap relative to the Fermi level could differ by up to $\sim 0.3$~eV.
The maximal ZT values were predicted by DFT  to vary from 1 to 3, for phonon thermal conductances in the range 
$10$ to $100$~pW/K. 

The high power factors we predict may be affected by inelastic transport channels, such as those arising from  the 
interaction of electrons with phonons. Here the current from inelastic processes may be significant, since the 
HOMO resonance is close to the anti-resonance. An incoming electron with an energy at the anti-resonance may 
emit a phonon with an energy that brings it close to the HOMO resonance and thereby bypass the anti-
resonance.
We have neglected such processes, but they appear to be import aspects for further studies of  thermoelectric 
properties of molecules relying on anti-resonances close to resonances. 
\begin{acknowledgements}
The authors thank Olov Karlstr{\"o}m for valuable discussions on the thermoelectric properties of molecules with 
destructive interference features. This work was supported by The European Union seventh Framework 
Programme (FP7/2007-2013) under the grant agreement no. 270369 ("ELFOS"). GCS and MS received funding 
from the European Research Council under the European Union's Seventh Framework Programme 
(FP7/2007-2013)/ ERC Grant agreement no. 258806.
\end{acknowledgements}
\appendix
\section{Transmission comparison}\label{sec.A1}
\begin{figure}[h!]
 \includegraphics[width=0.7\linewidth]{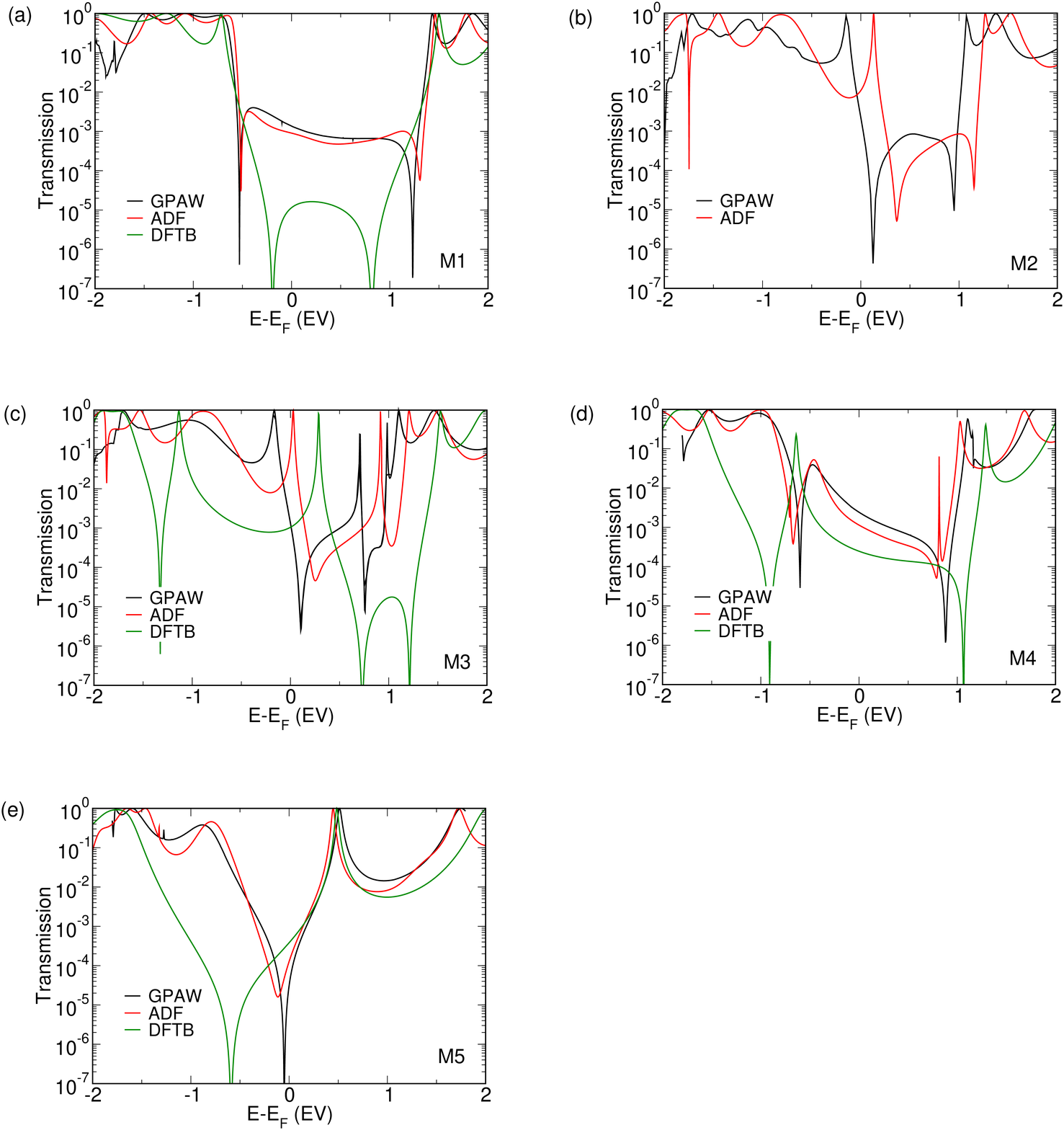}
  \caption{\label{fig.M1_to_M5_G_A_D_trans} Transmission calculated using GPAW (black line), ADF (red line), DFTB (green line).
  (a)-(e) corresponds to molecule M1-M5.}
\end{figure}

We compare in Figure \ref{fig.M1_to_M5_G_A_D_trans} 
the transmission calculated for molecule M1-M5 using GPAW, ADF and DFTB. We observe that the 
GPAW and ADF results are in overall good agreement, and that all three methods predicts the same
qualitative behaviour and trends in accordance with the electron withdrawing and donating character of the 
substituent side groups.

\section{Thermoelectrics}\label{sec.A2}
The transport coefficients relevant for thermoelectricity is written as 
\begin{eqnarray}
G&=&e^2L_0 \\ \label{eqn.g}
S&=&\frac{1}{eT}\frac{L_1}{L_0}\\ 
\kappa_{e}&=&\kappa_0-\frac{1}{T}\frac{L_1^2}{L_0}\\
                   &=&\kappa_0-TGS^2\\
\kappa_0&=&\frac{1}{T}L_2
\end{eqnarray}
 in terms of the function
\begin{equation}
L_n(\mu) = \frac{2}{h} \int d\varepsilon (\varepsilon-\mu)^n (-n_F'(\varepsilon))\tau(\varepsilon),
\label{eqn.L}
\end{equation}
where $n_F(\varepsilon)=(\exp((\varepsilon-\mu)/k_BT))+1)^{-1}$ is the Fermi-Dirac distribution function.
The thermal electronic conductance $\kappa_0$ is for zero chemical potential drop and related to the
thermal conductance at zero electric current by $\kappa_e=\kappa_0-TGS^2$. 

The dimensionless figure of merit which may be used to characterise the performance of a thermoelectric device is 
given by
\begin{eqnarray}
ZT&=&\frac{TGS^2}{\kappa_{e}+\kappa_{ph}}\nonumber\\
     &=&\frac{\kappa_0}{\kappa_{ph}} \left( \frac{1-\kappa_{e}/\kappa_0}{1+\kappa_{e}/\kappa_{ph}} \right)\\
     &\leq&\frac{\kappa_0}{\kappa_{ph}},
\end{eqnarray}
which we have rewritten in terms of the thermal conductances in the second line.
Since all thermal conductances are larger or equal to zero, we see immediately that
the expression in the parentheses is always smaller or equal to 1. This means that $\kappa_0/\kappa_{ph}$ is an upper bound 
to $ZT$ for a given transmission function and phonon thermal conductance. 
This upper bound, reached when $\kappa_e=0$, is sometimes referred to as the 
Mahan-Sofo (MS) bound \cite{ms_1996}. MS showed that the only a Dirac delta function as the transmission function gives
 $ZT=\kappa_0/\kappa_e$, i.e. the upper bound. However, when concerned with single molecule junctions, then not only is the integral of the transmission bounded but also the value at any energy. The transmission is usually smaller than $\sim1$ at the relevant energies within the HOMO-LUMO gap. This has the important consequence, that in the case of a infinitesimally narrow transmission resonance $\kappa_0\rightarrow 0$. As we shall see below, typically a finite width of the order of $\mathrm{k_B}T$ results in the highest figure of merit, even though this results in a $ZT$ below the MS bound.

\begin{figure}%[h!]
 \includegraphics[width=0.95\linewidth]{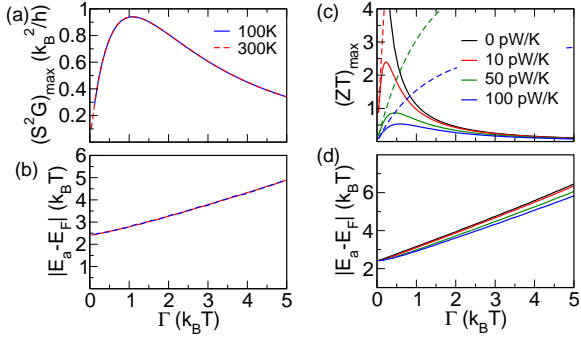}
  \caption{\label{fig.lorentzian_model} (a) Maximal 
  power factor as a function of the broadening $\Gamma$ for $T=100K$ (blue line) and $T=300K$ (red dashed line). 
  (b) Level position relative to the Fermi energy giving the maximal power factor in (a) for a given $\Gamma$. (c) Maximal figure of 
  merit as a function of $ \Gamma$ for $T=300$~K. The labels indicate the values of the contribution to the thermal conductance 
 from phonons. The dashed lines show the MS upper bound. 
  (d) The level position giving the maximal figure of merit in (c) for a given $\Gamma$.}
\end{figure}

\emph{Single level model}. We now consider a single level coupled to wide band leads for which 
the transmission takes the a simple Lorentzian form
\begin{equation}
\tau(\varepsilon)=\frac{\Gamma^2}{(\varepsilon-\varepsilon_a)^2+\Gamma^2},
\end{equation}
where $\Gamma$ gives the broadening and $\varepsilon_a$ is the level position.
We assume a temperature of $T=300$K unless otherwise stated.
In Figure \ref{fig.lorentzian_model}a we show the maximal power factor as a function of $\Gamma$. The level position relative to the 
Fermi level yielding the maximal power factor for a given $\Gamma$ is shown in Figure \ref{fig.lorentzian_model}b. We have used a 
temperature of 100K (blue line) and 300K (dashed red line) which gives the same results. Note that energy is in units of $\mathrm{k_B}T$.
The largest value for the maximal power factor is $\sim 0.9$ $\mathrm{k_B}^2/h = 258 fW/K^2$ obtained for $\Gamma\approx1.1$~$\mathrm{k_B}T$ and the level positioned about 2.9~$\mathrm{k_B}T$ away fro the Fermi level.

We show in Figure \ref{fig.lorentzian_model}c the maximal value of ZT, denoted as $\mathrm{(ZT)_{max}}$, as a function 
$\Gamma$ for different phonon thermal conductances, $\kappa_{ph}$, as indicated in the legends.
The dashed lines shows the MS upper bounds, 
and we see that only in the limit of small $\Gamma$ does the calculated 
$\mathrm{(ZT)_{max}}$ approach the MS bounds. However, in this limit ZT goes to zero. The largest value of 
$\mathrm{(ZT)_{max}}$ is obtained for a resonances with a with of about $\mathrm{k_B}T/2$ and 
the level energy positioned about 2.5~$\mathrm{k_B}T$ away from the Fermi level, see Figure \ref{fig.lorentzian_model}d. 
We obserbe that while $\mathrm{(ZT)_{max}}$
diverges in the limit of zero broadening for $\kappa_{ph}=0$~pW/K the highest possible value for a finite $\kappa_{ph}$,
quickly decreases.

\bibliography{myrefs}
\end{document}